\documentclass[%
 reprint,
 amsmath,amssymb,
 aps,
]{revtex4-1}

\usepackage{graphicx}% Include figure files
\usepackage{dcolumn}% Align table columns on decimal point
\usepackage{bm}% bold math
\begin{document}

\preprint{APS/123-QED}

\title{Features of the surface tension for the metal -- insulator
  boundary in the \\ vicinity of MI phase transition in the presence of external magnetic field.}% Force line breaks with \\
\thanks{e-mail: leoni@kiae.ru }%

\author{Leonid Dubovskii}
\affiliation{%
 NIC "Kurchatov Institute", 123182 Moscow.\\
 MFTI,  Dolgoprudnyi, 141700 Moscow region %\textbackslash\textbackslash
}%

%\affiliation{
%\affiliation{
 %Third institution, the second for Charlie Author
%\affiliation{%


\date{\today}% It is always \today, today,

\begin{abstract}
The self-consistent equations for MI phase transition are formulated. We assume
two order parameters which describe the phase transition. The first one is the density
distribution at MI boundary $\rho (\vec r)$. The second one is a two component complex vector
 in spin space $\Psi (\vec r)$. It determines electron density in metallic or semimetallic phase
  in the presence of external magnetic field. Two different components of the vector describe
  possible spin states of electrons inserted in the external magnetic field.
\par
The first order type MI phase transition determined by the variation of the density distribution
is considered by means of the gradient expansion of Cahn and Hillard type \cite{CahnHillard}.
The second order type transition of electron density beside MI boundary is described by
Ginzburg -- Landau expansion \cite{LandLif2}. The interaction between these two parameters is
assumed to be linear as a function of electron density with a coefficient which depends on
metallic density (cf. \cite{JinwuYe_Lubensky}). The obtained nonlinear equations are exactly
solved in the case of MI boundary in the presence of the parallel to the boundary or
perpendicular to it uniform magnetic field. The surface tension  $\Sigma _{mi}$ at the MI
boundary is calculated. It is shown that $\Sigma _{mi}$ is singular. In particular,
$\Sigma _{mi}\sim n^{3/2}$ as $ n\Rightarrow 0$ and $\Sigma _{mi}\sim (T-T_c (\vec h))^{3/2} .$
$T_c (\vec h)$ is the transition temperature in the presence of external magnetic field at MI phase transition.
\par
The singular behavior of $\Sigma _{mi}$ leads to an emphasized hysteresis at MI transition.
\begin{description}
\item[Usage]
to be published Pisma v ZhETF vol. 99, is. 1 (in Russian).
\end{description}
\end{abstract}

\pacs{Valid PACS appear here}% PACS, the Physics and Astronomy
\maketitle

%\tableofcontents

\section{Introduction.}
It is well known \cite{kik} that the values of surface tension of isolators are more than ten times
less than those of metals. The reason of this phenomenon is that the value of surface tension of a
metal is determined
by outflow of conduction electrons besides the boundary of the metal with vacuum at the distances
of the order of interplanar spacing of the metal \cite{Partenskiy}. In addition, the values of
surface tension of semimetals are of the same order as in common metals \cite{kik}.
It should be emphasized that the volume electron density in semimetals is of the order
of $10^{-5} $ from the density of the metals \cite{kik}.
\par
Any MI transition in a crystalline material, at any rate at zero temperature, must be a transition
from a situation in which bands overlap to a situation when they do not \cite{NFMott}. Small
band-crossing leads to a metallic state with a small number of electron carriers per atomic
cell like in semimetals.
%\subsection{A Subsection}
\section{The self-consistent equations of MI transition in external magnetic field. }
Our starting point is two-parameters equation of Ginzburg -- Landau type (cf. \cite{JinwuYe_Lubensky}). One of the order parameters of the transition is the density $\rho (\vec r)$ of metallic and isolator phases. The second order parameter is the electron carriers density which is described as a column with two complex components of electron wave function with different spin components:
\begin{equation}\label{gl1}
\Psi (\vec r)= \begin{pmatrix} \psi _{1}(\vec r)\\\psi _{2}(\vec r)
\end{pmatrix}\,\, , \,\,\,
\Psi ^{+}(\vec r)=(\psi _{1} ^* (\vec r), \psi _{2} ^* (\vec r))
\end{equation}
\par
Generally, $\psi _{1}(\vec r)$ and $\psi _{2}(\vec r)$ are two complex valued functions for
two components of spin. We put the following invariant scalar quantity:
\begin{equation}\label{gl2}
n(\vec r) =|\psi _{1}(\vec r)|^2 +|\psi _{2}(\vec r)|^2\equiv\left({\Psi ^{+}\Psi }\right)
\end{equation}
which is electron density in the metallic phase and becomes identically zero in the insulator phase at $T=0$.
\par
There are two additional invariant scalar quantities in the presence of magnetic field:
1) the energy of the magnetic field ${(\vec h)^2}/8\pi$ and 2) the convolution made by $\Psi ^{+}$ and
$\Psi $ of scalar product consisting from two vectors. One of the vectors is the magnetic field
vector $\vec h$. The second is Pauli vector matrix
$\vec{\sigma}=\vec i \sigma _x +\vec j \sigma _j +\vec k \sigma _z $.
\par
So, the Ginzburg-Landau functional may be assumed:
\begin{equation}\label{gln1}
\Phi=\Phi\{\rho ,\Psi , \Psi ^{+} \}=\int d\vec{r}\emph{F}(\rho , \Psi , \Psi ^{+} )\,  .
\end{equation}
\par
The functional for $\rho (\vec r)$ is expanded according to Cahn -- Hillard  \cite{CahnHillard},
see also \cite{LifKagan}, and the Ginzburg-Landau expansion \cite{LandLif2} is used for electron
density. The result is:
$$\emph{F}(\rho , \Psi , \Psi ^{+})=\varphi (\rho ) +\lambda (\rho )(\nabla\rho )^{2}+
\left(\alpha +g(\rho )\right)\left({\Psi ^{+}\Psi }\right)+$$
\begin{equation}\label{gl4}
+\frac{1}{2}\beta{\left({\Psi ^{+}\Psi }\right)}^{2}+\frac{\hbar ^2}{2m}
\left|\left(\nabla -\imath\frac{ e}{\hbar c}\vec A(r)\right)\Psi (\vec r)\right|^{2}+
\end{equation}
$$+{\vec h ^2}/8\pi +\mu\left(\Psi ^{+}\vec h\vec{\sigma}\Psi\right).$$
%\par
$g(\rho)$ describes interaction between two order parameters $\rho (\vec r)$ and $\Psi (\vec r)$,
$\alpha =a(T-T_c)\,$, $\vec{\sigma}$ -- Pauli matrixes. The last term in (\ref{gl4}) can be
represented as follows:
\begin{equation}\label{gln5}
\left(\Psi ^{+}\vec h\vec{\sigma}\Psi\right)=h_x(\psi _1 ^*\psi _2 +\psi _2 ^*\psi _1 )
-\imath h_y (\psi _1 ^*\psi _2 -\psi _2 ^*\psi _1 )+
\end{equation}
$$+h_z(\psi _1 ^*\psi _1 -\psi _2 ^*\psi _2 )\, .$$
\par
It is clear that (\ref{gln5}) is a real quantity. We vary the vectors in spin space $\Psi $,
$\Psi ^{+}$ and use (\ref{gln5}). The equations (\ref{gln1}--\ref{gl4}) become as follows:
$$-\frac{\hbar ^2}{2m}\left(\nabla -\imath\frac{ e}{\hbar c}\vec A(r)\right)^2 \Psi
+\left(\alpha +g(\rho )\right)\Psi+$$
\begin{equation}\label{gl7}
+\beta \left({\Psi ^{+}\Psi }\right) \Psi +\mu (\vec h \vec\sigma )\Psi =0
\end{equation}
As usual \cite{LandLif2}, the variation of components $\psi _1 ^*$,$\,\psi _2 ^*$,$\,\psi _1$ and
$\psi _2$ can be performed without mutual correlations.
\par
We vary $\rho (\vec r)$. The result is as follows:
\begin{equation}\label{gl8}
2\nabla\left(\lambda (\vec r)\nabla\rho (\vec r)\right)=\nabla\varphi (\rho)+\nabla g(\rho)
\left({\Psi ^{+}\Psi }\right)
\end{equation}
We vary the vector potential $\vec A$ as well. Using the equation $\vec h=rot \vec A$, we get
the equation for density of current identical with \cite{LandLif QM}:
%\rot \vec h=\frac{4\pi}{c}j(\vec r)\,\,\,\,\, ,\,\,\,\,
$$ j(\vec r)=\frac{\imath e\hbar}{2m}\left\{(\nabla\Psi ^+ ,\Psi)-(\Psi ^+ ,\nabla\Psi )\right\} -
 \frac{e^2}{mc}\vec A +$$
\begin{equation}\label{gl9}
+\mu\, rot (\Psi ^+ \vec\sigma \Psi )\,\, ,
\end{equation}
The Maxwell equations are also of importance:
\begin{equation}\label{gl10}
rot \vec h=\frac{4\pi}{c}\vec j(\vec r)\,\,\,\,\,  ,\,\,\,\,\,\,\,\,\,\,\,\,\,\, div\vec h=0
\end{equation}
So, the equations (\ref{gl7}) -- (\ref{gl10}) are the Ginzburg--Landau equations {\cite{LandLif2}
for our system.
\section{Exact solution of equations in the presence of magnetic field and surface tension at MI boundary . }
We start the analysis of the equations got for MI boundary. In this case all quantities
depend only on one space coordinate $x$ perpendicular to MI boundary.
The equation (\ref{gl8}) for $\rho (x)$ becomes  as follows:
\begin{equation}\label{r}
2\frac{d}{dx}\left(\lambda (\rho )\frac{d}{dx}\rho\right) =\varphi ' (\rho)+g' (\rho)\left({\Psi ^{+}\Psi }\right)
\end{equation}
Here prime means derivative on $x$. We consider the solution of the equation (\ref{r}) for $g\equiv 0$
when order parameters $\rho (x)$ and $\Psi (x)$ are not connected. The equation (\ref{r})
can be easily integrated:
\begin{equation}\label{gl61}
x=\int _{\rho _{min}} ^{\rho (x)} \lambda (\rho ^{'}) d\rho ^{'}
\left[\int _{\rho _{min}} ^{\rho ^{'}}\varphi ^{'}(\kappa)\lambda (\kappa)
d\kappa\right]^{-1/2}
\end{equation}
If the value $\lambda (\rho)$ does not depend on $\rho$, the equation (\ref{gl61}) becomes
as follows \cite{LifKagan}:
\begin{equation}\label{gl6}
x=\int _{0} ^{\rho (x)} d\rho \sqrt{\lambda/\varphi (\rho)}
\end{equation}
It is important to take into account the dependence  $\lambda (\rho)$ from $\rho$ as the
value $\lambda$ in metallic phase can differ significantly from that in isolator phase.
\par
Let dimension $[\varphi]=\varepsilon $ is the density of energy. In this case the value of $\lambda$ has
dimension $[\lambda]\simeq {\varepsilon}{l^{2} \rho ^{-2} }$ and proportional to $l^2$, where $l$  is
the length of the variation of density $\rho $. The value of surface tension $\Sigma _{\rho}$ of
MI interface reads as follows \cite{LifKagan}:
\begin{equation}\label{glq}
\Sigma _{\rho} =2\int _{\rho _{ins}} ^{\rho _{met}} d\rho\sqrt{\varphi (\rho)\lambda (\rho ) }
\end{equation}
Here we use only two first terms of (\ref{gl4}).
The value of the surface tension is $\Sigma _{\rho}\simeq\varepsilon l$. So, it has the dimension of the
energy attributed to unit of area. If $\lambda \Rightarrow 0$, the surface tension becomes zero.
If $l$ grows, the surface tension will grow as well.
\par
We assume the magnetic field is a function only of one coordinate $x$ perpendicular to MI
boundary. In addition, we assume the magnetic field has a component $h_z (x)$ in the plane of
the MI boundary and a component $h_x (x)$ perpendicular to it. So, the magnetic field and the
vector potential in Landau gauge \cite{LandLif QM} are as follows:
$\vec h=(h_x(x), 0, h_z(x))\, ;$
$$ \vec A(\vec r)=\left(0,\,\, \int _{-\infty} ^x h_z(x')dx'-zh_x(x), \,\ 0\right).$$
\par
For simplicity, we assume that MI transition is not accompanied by magnetic phase transition. In this
case, we take $h_x$ and $h_z$ do not depend on $x$ and the equation (\ref{gl7}) becomes:
\begin{equation}\label{gl7h1}
-\frac{\hbar ^2}{2m}\left(\frac{\partial ^2 \Psi}{\partial x^2}+\left(\frac{\partial}{\partial y}-\imath\frac{ e}{\hbar c}\left(x h_z-zh_x\right)\right)^2\Psi+\frac{\partial ^2 \Psi}{\partial z^2}\right)
+
\end{equation}
$$+\tilde{\alpha}\Psi+\beta \left({\Psi ^{+}\Psi }\right) \Psi +\mu\begin{pmatrix}h_z & h_x \\ h_x & -h_z \end{pmatrix}\Psi=0$$
$$\tilde{\alpha}=\left(\alpha +g(\rho )\right)$$
\par
We take $\vec A(\vec r)\equiv 0$. It means we neglect the orbital motion of electrons in magnetic
field in the metal. It is correct if electron free path $l_e$ is less than the radius of the electron
orbit in magnetic field $r_h$:
\begin{equation}\label{dingle}
l_e<2\pi r_h\, , \,\, r_h =\frac{cp_F}{eh}\,\, ,\,\,\,\,\, p_F - \text {Fermi momentum.}
\end{equation}
The magnetic field does not influence the motion of an electron between successive collisions.
It means the influence of magnetic field cannot be seen during electron mean free time. Really,
an electron moves straight line trajectory between successive collisions.
The inequality (\ref{dingle}) corresponds exactly to Dingle factor \cite{dingle} which determines
the absence of de Haas -- van Alpen oscillations due to impurity scattering. In addition, it is
assumed diffuse reflection at MI boundary.
\par
The other possibility when $\vec A(\vec r)\equiv 0$ corresponds the following inequality:
\begin{equation}\label{cor}
\xi _F<2\pi r_h,
\end{equation}
$\xi _F$ is the correlation length in the electron Fermi gas in metal.
Due to  \cite{LandLif} $\xi _F =\hbar v_F /\pi T$. The substitution of the last equation into (\ref{cor})
gives the inequality for the temperatures for which $\vec A(\vec r)$ can be neglected:
\begin{equation}\label{cor1}
 2\pi ^2 T> \hbar \Omega _h\,\,\, ,\,\,\,\,\,\,\, \Omega _h=\frac{eh}{mc}
\end{equation}
\par
The inequality (\ref{cor1}) coincides with the condition of the absence of de Haas -- van Alpen
effect due to the low value of the magnetic field as compared with
 the temperature \cite{LandLif} .
\par
If $\vec A(\vec r)\equiv 0$, we can write (\ref{gl7h1}) as a system of two connected nonlinear
equations for components $\Psi (x)$ (\ref{gl1}):
\par
$$-\frac{\hbar ^2}{2m}\frac{d^2 \psi _1}{dx^2}+\tilde{\alpha}\psi _1+\beta \left({{|\psi _1|}^2 +{|\psi _2|}^2}\right) \psi _1 +$$
\begin{equation}\label{glh3x}
+\mu (h_x\psi _2 + h_z\psi _1)=0
\end{equation}
%\begin{equation}\label{glh4x}
$$-\frac{\hbar ^2}{2m}\frac{d^2 \psi _2}{dx^2}+\tilde{\alpha}\psi _2+\beta \left({{|\psi _1|}^2 +{|\psi _2|}^2}\right) \psi _2  +$$
\begin{equation}\label{glh4x}
+\mu (h_x\psi _1 - h_z\psi _2)=0
\end{equation}
We try to find solutions of the system (\ref{glh3x}) -- (\ref{glh4x}) in the followig way:
\begin{equation}\label{glh5x}
\psi _2 (x)=q\psi _1 (x)\,\, , \,\,\,\,
\end{equation}
$q$ is a certain constant determined by $h_x$ and $h_z$.
The equations (\ref{glh3x}) and (\ref{glh4x}) coincide identically provided the
following equation is valid:
\begin{equation}\label{glh7}
h_x q + h_z=\frac{h_x}{q} - h_z
\end{equation}
It is clear that if $q=q_{\pm}$
\begin{equation}\label{glh7x}
 q_{+}=\frac{|h_x|}{\sqrt{h_x ^2 +h_z ^2}+h_z sgn{h_x}}\,\, and\,\,\,\,\, q_{-}=-\frac{1}{q_{+}}\, ,
\end{equation}
the equations  (\ref{glh3x}) and (\ref{glh4x}) coincide one with another.
We rewrite the equation (\ref{glh3x}) as follows:
\begin{equation}\label{glh5xy}
-\frac{\hbar ^2}{2m}\frac{d^2 \psi _1}{dx^2}+\tilde{\tilde{\alpha}}\psi _1+\tilde{\beta}(1+q^2) {|\psi _1|}^2 \psi _1 =0\,\, ;\,\,\,\,
\end{equation}
$$\tilde{\tilde{\alpha}}=\tilde{\alpha}+\mu (h_x q + h_z )\,\, ,\,\,\,\,\tilde{\beta}=\beta  (1+q^2)\,\, ;\,\,\,\, q=q_{\pm}\,\, .$$
The nonlinear equation (\ref{glh5xy}) has two uniform solutions. The first one $\psi _1 \equiv 0$
corresponds to the phase of isolator. The second one ${\psi _{1} ^0}$ corresponds to the
metallic phase with the following square modulus of the order parameter :
 \begin{equation}\label{glh7y}
 |{\psi _{1} ^0} | ^2 =-\frac{\tilde{\tilde{\alpha}}}{\tilde{\beta}(1+q^2)}  .
\end{equation}
The value $|{\psi _{2} ^0}| ^2 =q^2 |{\psi _{1} ^0}| ^2$ corresponds to the value $|{\psi _{1} ^0}| ^2\, . $
The sum of these quantities (see (\ref{gl2})) equals to the bulk electron density of metal $n_0$ :
$ |{\psi _{1} ^0}| ^2 +|{\psi _{2} ^0}| ^2 =n_0\,\, ,$
 $$\,\,\,\,\, |{\psi _{1} ^0}| ^2 =\frac{n_0}{(1+q^2)}\,\, ,
 \,\,\,\,\,|{\psi _{2} ^0}| ^2 =\frac{q^2 n_0}{(1+q^2)}\,\, .$$
We introduce the dimensionless function (\ref{glh5xy}) as follows:
\begin{equation}\label{glh9}
\psi _1 (x)={\psi _{1} ^0}f(x)
\end{equation}
The equation (\ref{glh5xy}) becomes:
\begin{equation}\label{glh10}
-\xi {^2}(T)\frac{d^2 f}{dx^2}-f+f^3  =0\,\, ,\,\,\,\,\,\,\,\,\,\,\,\,\,\,\, \xi (T)=\sqrt{\frac{\hbar ^2}{2m|\tilde{\tilde{\alpha}}|}}.
\end{equation}
The function $f(x)$ satisfys the following boundary conditions:
\par
$ f=0\,\,\, ,\,\,\,\,\,\,\,\,\,\,\,\,\,$ if $\,\,\,\, x=0$
\par
$ f\Rightarrow 1\,\,\, ,\,\,\,\,\,\,\,\,\,\,\,\,\,$ if$\,\,\,\, x\Rightarrow\infty$

The equation (\ref{glh10}) has the solution (cf. \cite{dG}):
\begin{equation}\label{gl16xx}
f(x)=\tanh\left[\frac{x}{\sqrt{2}\xi (T)}\right]\, .
\end{equation}
In the vicinity of phase transition $\tilde{\tilde{\alpha}}$ becomes zero.
$|{\psi _{1} ^0}|$  and $|{\psi _{2} ^0}|$ have square root singularity (see (\ref{glh7y})).
The function $f(x)$ changes its value at the coherence length $\xi (T)$. The coherence length goes
to infinity like square root singularity (\ref{glh10}).
\par
The calculation of the surface tension for MI boundary is in analogy with that for the
metal -- superconductor boundary (see \cite{dG})) and gives the following equation:
$$\Sigma _{\Psi}=\tilde{\tilde{\alpha}}{\psi _{1} ^0}^2(1+q^2)\int
_{0}^{\infty}
dx\left(\xi ^2 (T)\left(\frac{d{f }}{dx}\right)^2-\frac{1}{2}(1-f^2)^2\right)$$
We use the first integral of the equation (\ref{glh10}):
$$\xi {^2}(T)\left(\frac{df}{dx}\right)^2 =\frac{1}{2}(1-f^2)^2$$
and arrive at a goal:
$$\Sigma _{\Psi}=-\tilde{\tilde{\alpha}}{\psi _{1} ^0}^2(1+q^2)\int
_{0}^{\infty}
dx(1-f^2)^2=$$
\begin{equation}\label{st2xn1xx}
=-\tilde{\tilde{\alpha}}{\psi _{1} ^0}^2(1+q^2)\sqrt{2}\xi (T)=\beta\sqrt{2}n_0 ^2 \xi (T)
\end{equation}
It leads to $\Sigma _{mi}\sim n_0 ^{3/2}\,\,$ in the limit $ n_0\Rightarrow 0\, .$ In the
vicinity of the point MI transition in the magnetic field $T_c (\vec h)$ the value
$\Sigma _{mi}\sim (T-T_c (\vec h))^{3/2}\,\,\, .$
\par
So, the surface tension at MI boundary becomes as follows:
\begin{equation}\label{sigma}
\Sigma _{mi} =\Sigma _{\rho}+\Sigma _{\Psi}+\Sigma _{\rho\Psi}
\end{equation}
 $\Sigma _{\rho}$ is determined by (\ref{glq}), and $\Sigma _{\Psi}$ by (\ref{st2xn1xx}).
$\Sigma _{\rho\Psi}$ is determined by interaction between two order parameters $\rho (\vec r)$
and $\Psi (\vec r)$ and its value is of the order of (\ref{st2xn1xx}). However, it contains an
additional factor proportional to $g(\rho)$.
\section{Conclusions.}
The system of equations formulated in spin space describes MI phase transition in the external magnetic
field. The exact solution of this nonlinear system of equations is obtained in the external
uniform magnetic field. The magnetic field has an arbitrary direction relative to the MI boundary.
\par
It is shown that the surface tension for MI boundary has a singular behavior. The point of phase
transition behaves in different manner for the parallel to the boundary component of the magnetic field
and for perpendicular one. We assumed that the orbital motion in the magnetic field is suppressed
due to some type of scattering in the system as it usually takes place for MI phase transitions.
It can be scattering on impurities or temperature suppression of coherent motion.
\par
The singular behavior manifests itself for other thermodynamic and kinetic features of the system
close to MI transition. Probably, it take place in binary compounds with B20 structure like
MnSi, FeSi and FeGe, where recently uncommon properties where found. In particulary, in the FeSi
compound was found MI transition accompanied by anomalies of phonon spectra with a strong
temperature dependence \cite{Delaireetal}. It was discovered significant softening of phonon
spectrum by increase of temperature in the vicinity of MI phase transformation \cite{PPP}.
\par
I thank P.P.Parshin for that he attract my attention to anomalies of phonon spectrum in FeSi system
close to MI transition \cite{Delaireetal}, \cite{PPP}, and also for the possibility to get to know the
results of the paper \cite{PPP} before publication.
\par
The paper is fulfilled by partial support RFBR (grant 13-02-00469) and by partial support of the Department
of Education (grant 8364).
\par

\end{document}